\documentclass[prd,twocolumn]{revtex4}

\pdfoutput=1

\usepackage{graphicx}
\usepackage{amssymb,amsmath,bm}  
\usepackage[usenames,dvipsnames]{xcolor}
\usepackage{hyperref}
\hypersetup{
    colorlinks = true,
    citecolor = {MidnightBlue},
    linkcolor = {BrickRed},
    urlcolor = {BrickRed}
}

\newcommand{\uKam}{\mu\text{K-arcmin}}

\newcommand{\bl}{{\bm \ell}}
\newcommand{\bx}{{\bm x}}

\begin{document}
\title{Calibrating Cluster Number Counts with CMB lensing}
\author{ Thibaut Louis$^1$ and David Alonso$^2$}
\affiliation{$^{1}$UPMC Univ Paris 06, UMR7095, Institut d'Astrophysique de Paris, F-75014, Paris, France\\
             $^{2}$ University of Oxford, Denys Wilkinson Building, Keble Road, Oxford OX1 3RH, UK
             }

\begin{abstract}
  CMB Stage-4 experiments will reduce the uncertainties on the gravitational lensing potential by
  an order of magnitude compared to current measurements, and will also produce a Sunyaev-Zel'dovich
  (SZ) cluster catalog containing $\sim10^{5}$ objects, two orders of magnitudes higher than what
  is currently available. In this paper we propose to combine these two observables and show that
  it is possible to calibrate the masses of the full Stage-4 cluster catalog internally
  owing to the high signal to noise measurement of the CMB lensing convergence field. We find that
  a CMB Stage-4 experiment will constrain the hydrostatic bias parameter to sub-percent accuracy.
  We also show constraints on a non parametric $Y-M$ relationship which could be used to study
  its evolution with mass and redshift. Finally we present a joint likelihood for thermal
  SZ (tSZ) flux and mass measurements, and show that it could lead to a $\sim5\sigma$ detection of
  the lower limit on the sum of the neutrino masses in the normal hierarchy 
  ($\sum m_{\nu}=60 \textrm{meV}$) once combined with measurements of the primordial
  CMB and CMB lensing power spectra.
\end{abstract}

  \date{\today}
  \maketitle

\section{Introduction}\label{sec:intro}

The number of galaxy clusters as a function of mass and redshift is a prediction
of the LCDM model, and by accurately reconstructing the cluster mass function we can
put constraints on cosmological parameters such as the matter density $\Omega_m$,
the sum on the neutrinos masses $\sum m_{\nu}$ and the normalisation of the linear
matter power spectrum $\sigma_{8}$.  This observable has recently gained a vivid
interest following the publication of the Planck cluster catalog ($\approx 10^{3}$
clusters) and the slight discrepancy between cosmological parameters inferred from
the primary CMB and from the distribution of cluster masses
\cite{2015arXiv150201598P,2015arXiv150201597P}. 

A classical method to estimate cluster masses is to use measurements of the
Compton-$y$ parameter, a measurement of the integrated flux of the thermal
Sunyaev-Zel'dovich effect at the position of the cluster.
The size of the effect is proportional to the total thermal energy of the cluster
gas and is therefore correlated with cluster mass \cite{2008ApJ...675..106B,
2012ApJ...754..119M,2013ApJ...772...25S}. This $Y-M$ scaling relation has traditionally
been determined empirically from X-ray observations of clusters \cite{2010A&A...517A..92A,
2014A&A...571A..20P}. However, X-ray-inferred cluster masses rely on the assumption
that clusters have reached hydrostatic equilibrium, and departure from this equilibrium
can bias the estimated cluster masses.  Physical phenomena causing this departure
include bulk motions in the gas or non-thermal sources of pressure (such as magnetic
fields or cosmic rays). Numerical simulations have shown that this can lead to
an underestimation of the true cluster masses by 10 to 15 $\%$
\cite{2008A&A...491...71P,2007ApJ...668....1N,2010A&A...514A..93M}. Moreover,
instrumental systematics in the X-ray analysis could propagate into the
cosmological results \cite{2012MNRAS.426.2046A}, and  an independent 
method for calibrating the scaling relation is extremely valuable.

Following the first detections of CMB lensing by clusters \cite{2015PhRvL.114o1302M,
2015ApJ...806..247B}, a new method for self-calibrating the cluster masses using
measurements of the lensing convergence at the cluster positions has recently been
proposed \cite{2015A&A...578A..21M}. It has been demonstrated on simulations and
successfully applied to Planck data resulting in a $5 \sigma$ measurement of the
hydrostatic bias parameter \cite{2015arXiv150201597P}. The aim of this work is to
discuss extensions of this method in the era of CMB Stage-4 (S4), a next-generation
CMB experiment that will achieve a cosmic-variance-limited reconstruction of the
convergence field up to multipoles $\ell \sim 1000$.

This paper is structured as follows. In Section \ref{sec:method} we describe our
cluster lensing model as well as a maximum likelihood estimator for the cluster
masses from a lensing convergence map. We also discuss the impact of possible
foregrounds contamination and atmospheric noise. In
Section \ref{sec:calibration} we propose two parametric methods to calibrate
the $Y-M$ relationship using cluster lensing. First we forecast constraints on the
hydrostatic parameter following the method proposed in \cite{2015A&A...578A..21M},
and then extend the formalism and forecast constraints on more general scaling
relations, including a non-parametric model that can be used to study the mass
and redshift dependence freely. In Section \ref{sec:cosmopar} we study how the
availability of joint tSZ and lensing mass measurements improves the cosmological
constraints achievable by a cluster survey carried out with S4 by consistently
accounting for the uncertainties in the $Y-M$ scaling relation. We summarise
our main conclusions in Section \ref{sec:conclusion}.

Throughout this paper we adopt a fiducial cosmology with $\Omega_{m}=0.315$,
$\Omega_{b}=0.049$, $\Omega_{\Lambda}= 0.685$, $H_{0}=67\,
{\textrm{km s}^{-1}\textrm{Mpc}^{-1}}$, $A_s=2.2\times10^{-9}$, $n_s=0.96$
and $\tau=0.06$, compatible with \cite{2015arXiv150201589P}.
We will use cluster masses $M_{500}$ defined as the mass measured within
a radius $R_{500}$ that encloses a mean density $500$ times larger than the
critical density at the cluster redshift. We will also estimate the number
density of haloes as a function of mass using the measurements of the
mass function by \cite{2008ApJ...688..709T}.

\section{Cluster lensing model}\label{sec:method}

The potential of CMB lensing to determine cluster masses has been
long recognised \cite{2000ApJ...538...57S,1999PhRvD..59l3507Z,2004ApJ...616....8H,
2004NewA...10....1V}. In this section, we follow \cite{2015A&A...578A..21M} and
describe a maximum likelihood estimator for clusters masses based on measurements
of the lensing convergence map. We then discuss the noise properties of CMB
Stage-4, and the possible contamination of the lensing field reconstructed
from temperature data due to atmospheric noise and foregrounds.

\subsection{Matched filter estimate of cluster masses}\label{ssec:maxLike}
We start by modelling the cluster mass distribution using a NFW profile
\cite{1996ApJ...462..563N}
\begin{equation}
\rho(x)= \frac{\rho_{0}}{(c_{500}x_r)(1+(c_{500}x_r))^{2}}
\end{equation}
where $\rho_{0}$ is the central mass density, $x$ is a dimensionless radial
variable $x_r=r/R_{500}$ and $c_{500}$ is the concentration parameter. In what
follows we will use a constant concentration $c_{500}=1.18$. The halo lensing
convergence can be related to the cluster surface mass density via
$\kappa(\bx)=\Sigma(\bx)/\Sigma_{\rm crit}$, where
\begin{align}
  \Sigma(\bx)\equiv\int_{-\infty}^\infty dl\,\rho(l,\bx),&
  \hspace{6pt}
  \Sigma_{\rm crit}\equiv\frac{c^2d_S}{4\pi G\,d_L\,d_{LS}}.  
\end{align}
Here $d_L$, $d_S$, and $d_{LS}$ are the angular diameter distances to 
the lens (the cluster), the source (the CMB) and the angular diameter distance
between lens and source.

Consider now a patch centered on a galaxy cluster of mass $M_{500}$, our data
model for the convergence map at the cluster position is
\begin{equation}
\kappa(\bx)= U_{\kappa}(\bx) \kappa_{5 \theta_{500}} + n_\kappa(\bx)
\end{equation}
with $\theta_{500}\equiv R_{500}/d_L(z)$.
$\kappa_{5 \theta_{500}}$ is the convergence integrated on a disc of radius
$5\theta_{500}$, which can be simply related to the cluster column mass in
a cylinder of radius $5R_{500}$,
$M_{5R_{500}}=d_L^2\Sigma_{\textrm{crit}}\kappa_{5\theta_{500}}$
\cite{2015arXiv150201597P}. $U_{\kappa}(\bx)$ is the normalised cluster
convergence spatial template, defined as
\begin{equation}
U_{\kappa}(\bx)=
\frac{\kappa^{\textrm{true}}(\bx)}{\kappa_{5 \theta_{500}}}=
\left[2\pi\int^{5\theta_{500}}_{0}dx\,x\,\Sigma(x)\right]^{-1}\Sigma(\bx),
\end{equation}
and $n_\kappa(\bx)$ is a stochastic noise term, containing contributions both
from the lensing reconstruction noise and from the lensing signal arising
from other structures along the line of sight. The latter component is
modelled here as a Gaussian field with the standard $\Lambda$CDM power spectrum
$C_{\kappa \kappa}(\ell)$. A minimum variance estimator for
$\kappa_{5 \theta_{500}}$ can be obtained as
\begin{equation}
\hat{\kappa}_{5\theta_{500}}=\sigma^2(\hat{\kappa}_{5\theta_{500}})
\int d\bl U^T_{\kappa}(\bl)N_{\kappa\kappa}^{-1}(\bl)\kappa(\bl),
\end{equation}
where $N_{\kappa\kappa}$ is the noise power spectrum,
$\langle n_\kappa(\bl)n_\kappa^*(\bl')\rangle=\delta(\bl-\bl')\,
N_{\kappa\kappa}(\bl)$, and the variance on $\tilde{\kappa}_{5\theta_{500}}$ is given by
\begin{equation}
\sigma^{-2}(\hat{\kappa}_{5\theta_{500}})=\int d\bl
U^T_{\kappa}(\bl)N_{\kappa\kappa}^{-1}(\bl)U_{\kappa}(\bl).
\end{equation}
Note that $\sigma^2(\kappa_{5\theta_{500}})$ is linearly related to the
variance on the mass measurement
\begin{equation}
\sigma^{2}(\hat{M}_{5 R_{500}})=\left[d_L^{2}\,\Sigma_{\textrm{crit}}\right]^{2}
\sigma^{2}(\hat{\kappa}_{5\theta_{500}} ).
\label{eq:sigmaM}
\end{equation}

\subsection{Noise on the convergence maps}\label{ssec:noise}
\begin{table}
  \centering{
  \renewcommand*{\arraystretch}{1.2}
  \begin{tabular}{|c|c|c|}
  \hline
  Frequency & Noise RMS & Beam FWHM \\
  (GHz) & ($\uKam$) & (arcmin) \\
  \hline
  ~28 & 9.8 & 14.0  \\
  ~41 & 8.9 & 10.0  \\
  ~90 & 1.0 & ~5.0   \\
  150 & 0.9 & ~2.8 \\
  230 & 3.1 & ~2.0 \\
  \hline
  \end{tabular}}
  \caption{Specifications for a S4 CMB experiment. The frequency bands were chosen to 
           lie on the main atmospheric windows, and the noise levels were designed to
           yield a map-level rms noise of $\sim1\uKam$ after foreground cleaning. The 
           highest and lowest frequency channels could be used to clean dust, cosmic
           infrared background and synchrotron contamination. In this analysis, we
           will only include the 41, 90 and 150 GHz channels.}
  \label{tab:cmbexp}
\end{table}
    
Lensing generates off diagonal correlations between different CMB multipoles. The
standard approach to reconstruct the convergence field is to form quadratic
estimators from the lensed CMB maps $\langle \tilde{X}({\bl}) \tilde{Y}^{*}({\bl'})
\rangle \propto \kappa_{XY}(\bl+\bl')$ \cite{2002ApJ...574..566H,2006PhR...429....1L}.
Given a set of experimental specifications (noise level and beam size) it is
possible to predict the expected reconstruction noise on the convergence
field. We will study two different reconstruction schemes:
\begin{itemize}
  \item Full reconstruction where the five quadratic estimators $\kappa_{TT},
        \kappa_{TE},\kappa_{TB},\kappa_{EE},\kappa_{EB}$  are used to reconstruct
        the convergence field.
  \item Polarisation-only reconstruction where we drop all estimators based on
        the temperature map and use only the polarisation based estimators
        $\kappa_{EE},\kappa_{EB}$.
\end{itemize}
\begin{figure}
  \centering
  \includegraphics[width=0.5\textwidth]{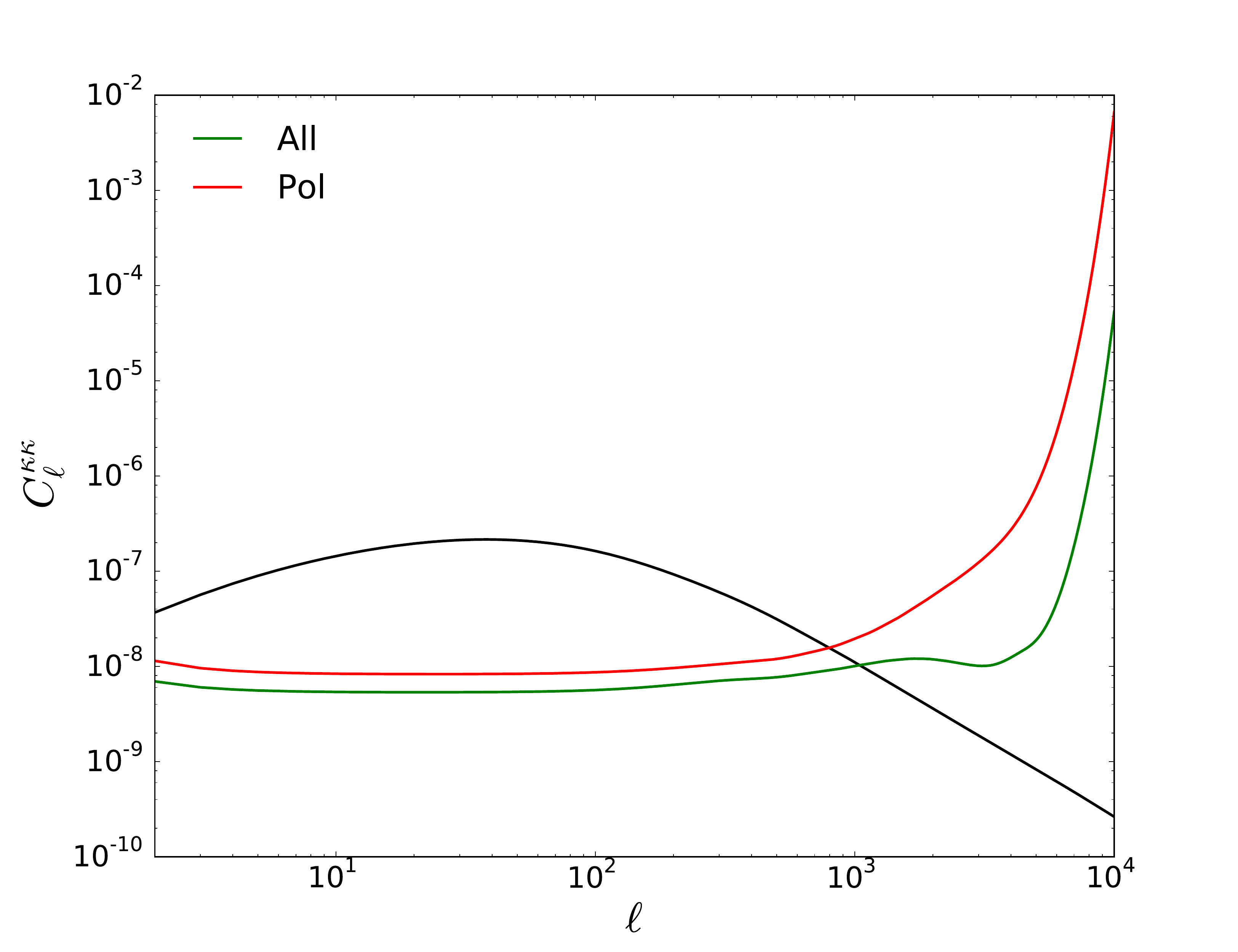}
  \caption{Lensing noise for the two minimum variance combinations highlighted
           in the text, containing all two-point combinations (green) and
           polarisation-only combinations (red). The CMB lensing convergence
           power spectrum is shown as a black solid line. The noise curves
           correspond to the survey specifications in Table \ref{tab:cmbexp}.}
  \label{fig:lnoise}
\end{figure}
The reason for studying these two different cases is twofold. First, on the
smallest angular scales, the level of foreground contamination is higher in
temperature than polarisation \cite{2014JCAP...10..007N,2015ApJ...805...36C}.
When partial foreground cleaning is possible owing to the multifrequency coverage
of S4, the leftover systematic effect due to incorrect foreground modeling could
hamper the reconstruction. Moreover, the kinematic SZ (kSZ) effect, present mostly in
temperature, has the same black-body spectrum as the primordial CMB and can not
be subtracted using multifrequency observation. At the level of precision targeted
by CMB S4 experiments this could lead to significant biases in the convergence
map. The second reason for ignoring the temperature-based estimators is
atmospheric noise. Since the likely implementation of S4 will be in the form of
a set of ground-based facilities, the largest angular scales will suffer from
contamination due to atmospheric emission (see for example Figure 2 of
\cite{2014JCAP...04..014D}). While the exact level of this atmospheric
contamination will depend on the geographical location of S4, as well as on its scanning
strategy, we choose to follow a conservative approach presenting as a baseline
the results for polarisation-only estimators, where atmospheric contamination is
smaller and can be mitigated through the
use of half-wave plates \cite{2016JLTP..184..534S}. Considering polarisation-only
estimators further allows us to fully decouple the measurements of the cluster
convergence and thermal Sunyaev-Zel'dovich effect.

The experiment specifications of S4 assumed here are reported in Table
\ref{tab:cmbexp}.
The frequency channels were chosen to lie on the atmospheric
windows, the relative noise levels were defined assuming template foreground
cleaning from synchrotron at low frequencies (spectral index $\beta_s=-3$) and
from thermal dust emission at high frequencies (spectral index $\beta_d=1.5$).
The absolute noise scale was chosen to yield a map-level RMS noise of $1\uKam$,
and the beam widths correspond to a 3m aperture telescope. We further assume
a total surveyed area of $f_{\rm sky}=0.4$. The lensing noise corresponding
to the two minimum-variance combinations (with and without temperature information)
for these specifications are shown in Fig.~\ref{fig:lnoise}, and were computed
using {\tt quicklens}\footnote{\url{https://github.com/dhanson/quicklens}}.

\section{Calibration of the $Y-M$ relationship}\label{sec:calibration}
In this section, we discuss the tight scaling relationship between the integrated
tSZ flux $Y$ emitted by a cluster and its mass, and how to calibrate this relationship
using measurements of cluster gravitational lensing. We start by discussing our
method to estimate the statistics of the cluster sample achievable with S4, which
determines the accuracy with which this relationship can be constrained. We then
consider the empirical $Y-M$ scaling relation
\begin{equation}\label{eq:yscaling}
  Y_{500}=\tilde{A}_{Y}
  \left[\frac{d_A(z)}{100\,{\rm Mpc}/h}\right]^{-2}
  \left[\frac{(1-b)M_{500}}{1.5\times10^{14}\,M_\odot/h}\right]^{\alpha_{Y}}
  E^{3/2}(z),
\end{equation}
where $d_A$ is the angular diameter distance to the cluster, $\alpha_{Y}=1.79$,
$\tilde{A}_{Y}=5.0\times10^{-10} {\rm sr}^2$, and $E(z)\equiv H(z)/H_0$
\cite{2015arXiv150201597P}. We will present forecasts for constraints on the
hydrostatic parameter $b$, on a generic power-law model and on a non-parametric
mass- and redshift-dependent $Y-M$ relationship.
 
\subsection{Cluster detection}\label{ssec:tSZ}
\begin{figure}
  \centering
  \includegraphics[width=0.55\textwidth]{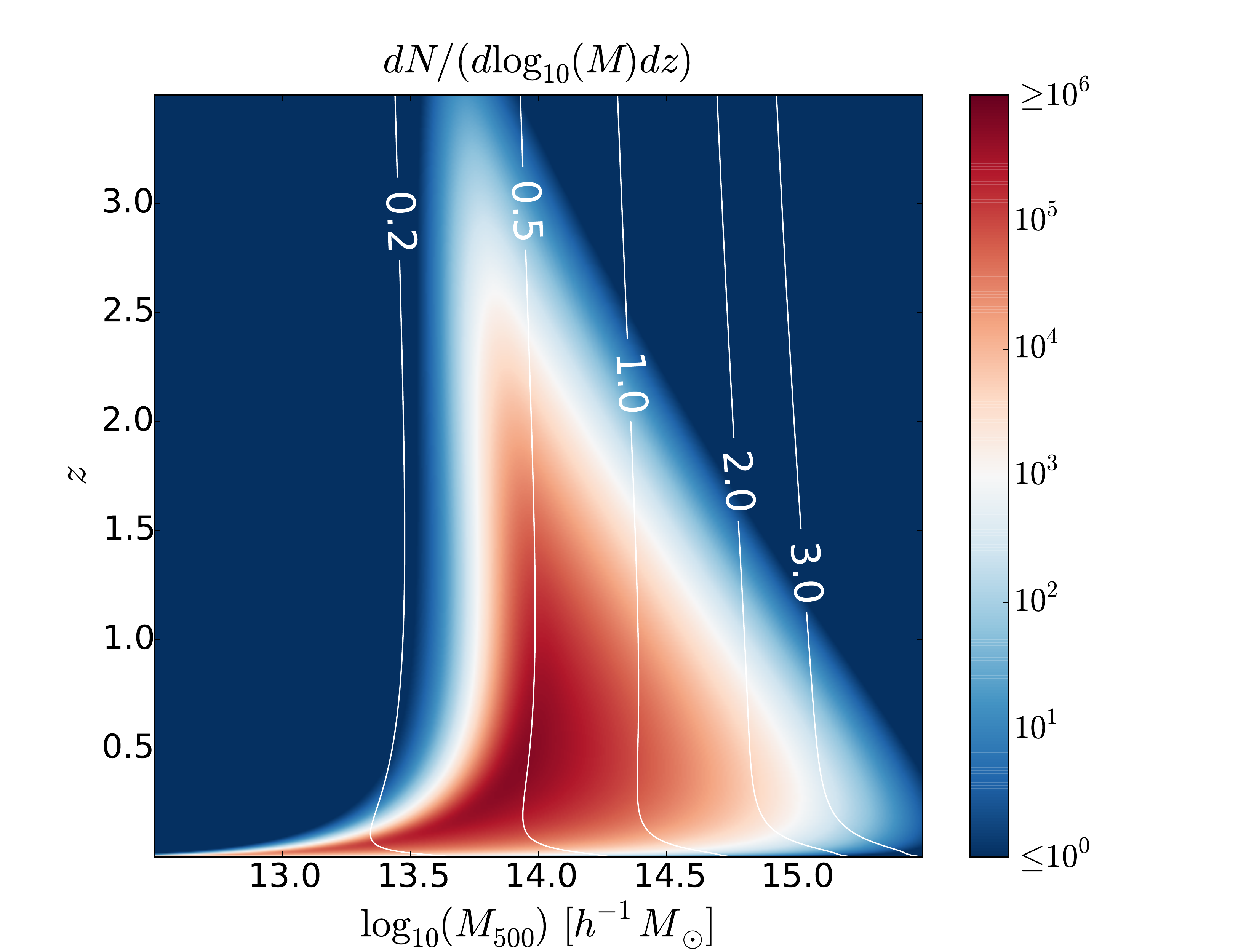}
  \caption{Mass and redshift distribution of  tSZ detected clusters for
           a CMB S4 experiment, the contour lines correspond to the signal
           to noise on the mass measurement of each individual cluster
           using CMB lensing estimated from polarisation data.}
  \label{fig:signaltonoise}
\end{figure}    
The tSZ effect arises due to inverse Compton scattering of CMB photons with
the hot electron gas inside clusters, it results in secondary contributions
to the CMB anisotropies \citep{1980ARA&A..18..537S}
\begin{align}\nonumber
  \left . \frac{\Delta {\rm T}}{{\rm T}} \right |_{\rm tSZ}(\nu,\bx)&=
  f_{\rm tSZ}(\nu)\frac{\sigma_T}{m_ec^2}\int P_e(l,\bx)\, dl\\\label{eq:tsz}
  &\equiv f_{\rm tSZ}(\nu) \, y(\bx),
\end{align}
where  $P_e = k_{\rm B} n_e T_e$ is the electron pressure, $\sigma_T$ is the
Thomson scattering cross-section, and where we have defined the dimensionless
Compton-$y$ parameter, $y(\bx)$.

The spectral signature of the tSZ effect  $f_{\rm tSZ}(\nu)$ allow us to
separate it from other types of emission in microwave frequency bands.
We start by defining the cluster detection efficiency 
\begin{align}
  \tilde{\chi}(M_{500},z)=&\int d(\ln Y^{\rm true}_{500}) \int_{q\sigma_N}^\infty
  dY_{500}^{\rm obs} \label{eq:chi_sz1}\\
  &~~~~~ P_{\rm SZ}(\ln Y^{\rm true}_{500}|M_{500},z)\,
  P_{\rm det}(Y_{500}^{\rm obs}|Y_{500}^{\rm true}), \nonumber
\end{align}
where $P_{\rm det}$ is the probability of obtaining a measurement $Y_{500}^{\rm obs}$
for a true integrated tSZ flux $Y_{500}^{\rm true}$, and
$P_{\rm SZ}$ is the distribution of integrated tSZ fluxes for clusters of mass
$M_{500}$ at redshift $z$, which accounts for the intrinsic scatter in the $Y-M$
relation. We have defined the tSZ flux $Y_{500}$ as the normalised integral of the
cluster pressure profile on a sphere of radius $R_{500}$. We refer the reader
to Appendix A of \cite{Alonso:2016jpy} for a precise definition of $Y_{500}$ as
well as the formalism used to model the tSZ catalog achievable by S4.

The detection efficiency is determined by the noise in the $Y$-measurements.
For this, and following \cite{Alonso:2016jpy}, we used a matched-filter approach,
which allows us to obtain optimal uncertainties simulaneously marginalised over
the amplitude of the kSZ effect. We find that S4 will be able to produce a catalog
with $\sim2\times10^{5}$ clusters for a tSZ signal-to-noise threshold $q_Y>6$,
two orders of magnitude above what is available today. With such a large number of
sources, a number of them will inevitably overlap on the sky, complicating the accurate
measurement of their individual tSZ or lensing signatures. In order to determine the
magnitude of this problem we start by defining the radius $\theta_{\rm blend}$ as the
aperture containing 95\% of the beam-convolved pressure profile for the typical cluster
size in the catalog. We then use this value to estimate the fraction of clusters that
overlap with other sources within $\theta_{\rm blend}$ to be $f_{\rm blend}\sim0.25$.
In the rest of this analysis we therefore remove 25\% of the cluster sample. Note that
this is a conservative approach, since blended sources could in principle be disentangled
given an accurate model of their profiles. The mass and redshift distribution of the
resulting catalog, as well as the signal-to-noise on individual cluster masses
measured with CMB lensing are shown in Fig.~\ref{fig:signaltonoise}. The $S/N$
increases with cluster mass but is not strongly dependent on its redshift. 

In the following, we will assume that each cluster in the catalog
experiment has a counterpart in an overlapping spectroscopic or photometric
galaxy survey, and that the errors on the redshift of individual clusters
can be neglected. CMB S4 is currently designed to have a full overlap with
LSST \cite{2009arXiv0912.0201L} and 4MOST \cite{2014SPIE.9147E..0MD}, and
while redshift uncertainties might play a role in the calibration of the
$Y-M$ relationship, they are always subdominant compared to uncertainties
in the cluster mass inferred from gravitational lensing.

\subsection{Hydrostatic bias}\label{ssec:hb}
As was mentioned in Section \ref{sec:intro}, cluster masses inferred from X-ray
observations can be biased due to the assumption of hydrostatic equilibrium (HE).
This effect can be taken into account by introducing the hydrostatic bias
parameter: $M_{Y}=M^{\rm HE}_{500}=(1-b)M_{500}$.
Gravitational lensing of the CMB provides an unbiased way of measuring
cluster masses $M_{L}= M_{500}$ and can be used to put a prior on $b$ \cite{2015arXiv150201597P}. 

\begin{figure}
\centering
  \includegraphics[width=0.55\textwidth]{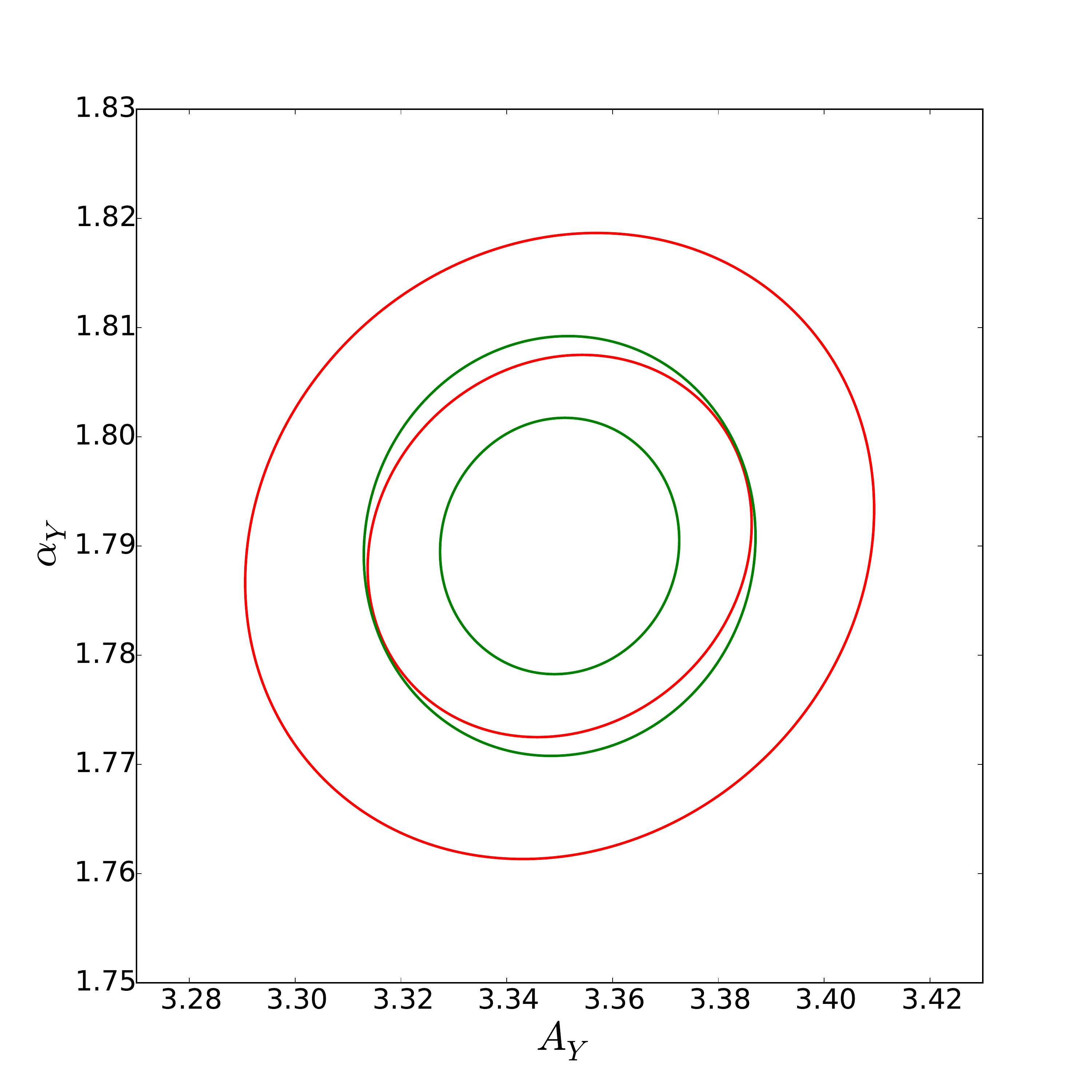}
  \caption{Constraints on the parameters of the $Y-M$
           scaling relationship for CMB S4 experiment. The 1 and 2$\sigma$
           contours are displayed. In green we show the result for the
           full reconstruction of the convergence field and in red
           the result based on polarisation data only.}
  \label{fig:scaling}
\end{figure}
The two mass measurements uncertainties are
\begin{equation}
\frac{\sigma^{2}(M_{Y})}{(M_{Y})^{2}}=
\alpha^{2}_Y\frac{\sigma^{2}({Y}_{500})}{{Y}^{2}_{500}}, \hspace{6pt}
\frac{\sigma^{2}(M_{L})}{(M_{L})^{2}}=
\frac{\sigma^{2}( \kappa_{5 \theta_{500}}) }{\kappa^{2}_{5 \theta_{500}}}
\end{equation}
The relative error on $\epsilon=(1-b)$ for a single cluster is then given by:
\begin{align}\nonumber
\frac{\sigma^{2}(\epsilon)}{\epsilon^{2}}&=\frac{\sigma^{2}(M_{Y})}{(M_{Y})^{2}}+
\frac{\sigma^{2}(M_{L})}{(M_{L})^{2}}\\&=\alpha_Y^{2}\frac{\sigma^{2}({Y}_{500})}
{{Y}^{2}_{500}}+\frac{\sigma^{2}(\kappa_{5\theta_{500}})}{\kappa^{2}_{5\theta_{500}}}.
\end{align}
Assuming independent estimates for each cluster, the uncertainty for the total
cluster sample is:
\begin{align}
\left[\frac{\sigma(\epsilon)}{\epsilon}\right]^{-2}_{\rm tot}&=
\sum^{N_{c}}_{i=1} \left[\frac{\sigma_{i}(\epsilon)}{\epsilon}\right]^{-2} \\
&=\int dz\frac{dV}{dz}
\int dM\frac{n(M,z)\tilde{\chi}(M,z)}{{\sigma^{2}(\epsilon)\epsilon^{-2}}(M,z)},
\nonumber
\end{align}
where $dV/dz\equiv 4\pi\,f_{\rm sky}\,c\,r^2(z)/H(z)$ is the derivative of
the comoving volume as a function of redshift, and $n(M,z)$ is the halo mass
function (comoving number density of haloes in a differential bin of mass).

The relative uncertainties corresponding to the two different lensing
reconstruction schemes described Section \ref{ssec:noise} are
\begin{align}
\frac{\sigma(\epsilon)}{\epsilon}&=
 2.51 \times 10^{-3} \hspace{0.2cm}(\textrm{Temp.+Pol.}) \\
\frac{\sigma(\epsilon)}{\epsilon}&=
3.97  \times 10^{-3} \hspace{0.2cm}(\textrm{Pol. only}).
\end{align}
Cluster lensing of the CMB therefore allows us to constrain the hydrostatic bias
parameter to sub-percent accuracy. This is useful, not only as a strong
prior when extracting cosmological constraints from cluster number counts,
but also as a source of information on cluster gas physics. In this
computation we have fixed all parameters of the scaling relationship,
assuming that X-ray observations could be used to put strong priors on them.
In the following we will show that lensing measurements by
CMB S4 could be used to constrain these parameters directly.

\subsection{Scaling relation calibration}\label{ssec:scaling}
In this section we explore the possibility of using CMB
lensing alone to constrain the $Y-M$ relationship, thus bypassing 
the need for X-ray follow-up observations and the uncertainties
associated with the hydrostatic bias. We start by parametrising
the $Y-M$ relationship as a scaling law
\begin{equation}
\frac{Y_{500}}{10^{-10}\,{\rm srad}}= E^{2/3}(z)
\left(\frac{100\,{\rm Mpc}/h}{d_A(z)}\right)^2
A_{Y}\left(\frac{M_{500}}{M_*}\right)^{\alpha_{Y}},
\end{equation}
with pivot scale $M_*=1.5\times10^{14}M_\odot\,h^{-1}$.
\begin{figure*}
  \centering
  \includegraphics[width=0.95\textwidth]{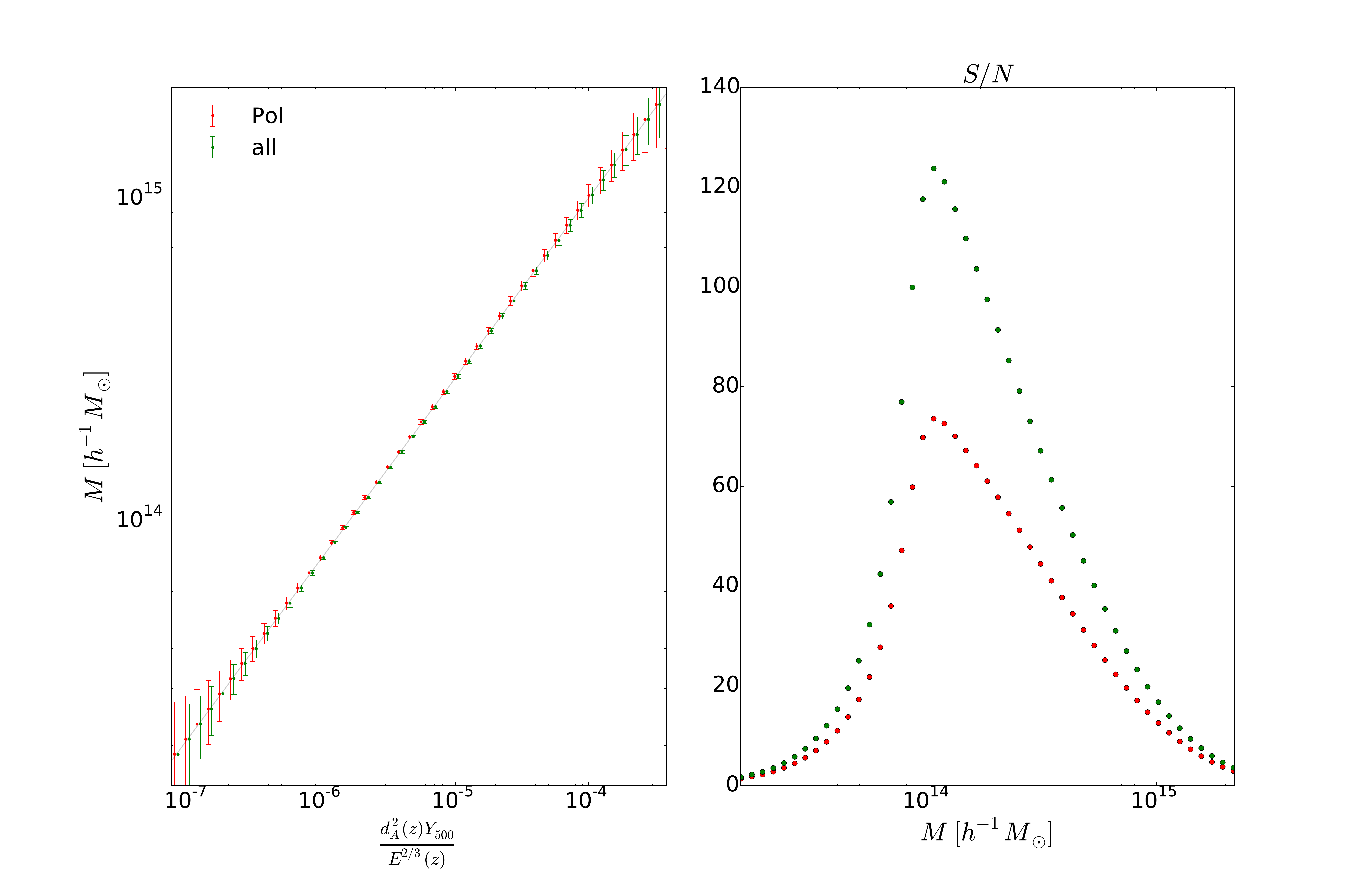}
  \caption{Non-parametric reconstruction of the ${ \cal Y}-M$ relationship
           using estimate of the convergence field from the polarised CMB.
           The left panel shows the error bars on the scaling relation, the
           right panel shows the S/N for each of the 44 mass bins. Note that
           the S/N will depend on the exact choice of model that is assumed for
           the relationship. Results are shown in green for the minimum variance 
           lensing reconstruction and in red for reconstruction using only
           polarisation-based estimators.}
  \label{fig:blind}
\end{figure*}
\begin{figure}
  \centering
  \includegraphics[width=0.5\textwidth]{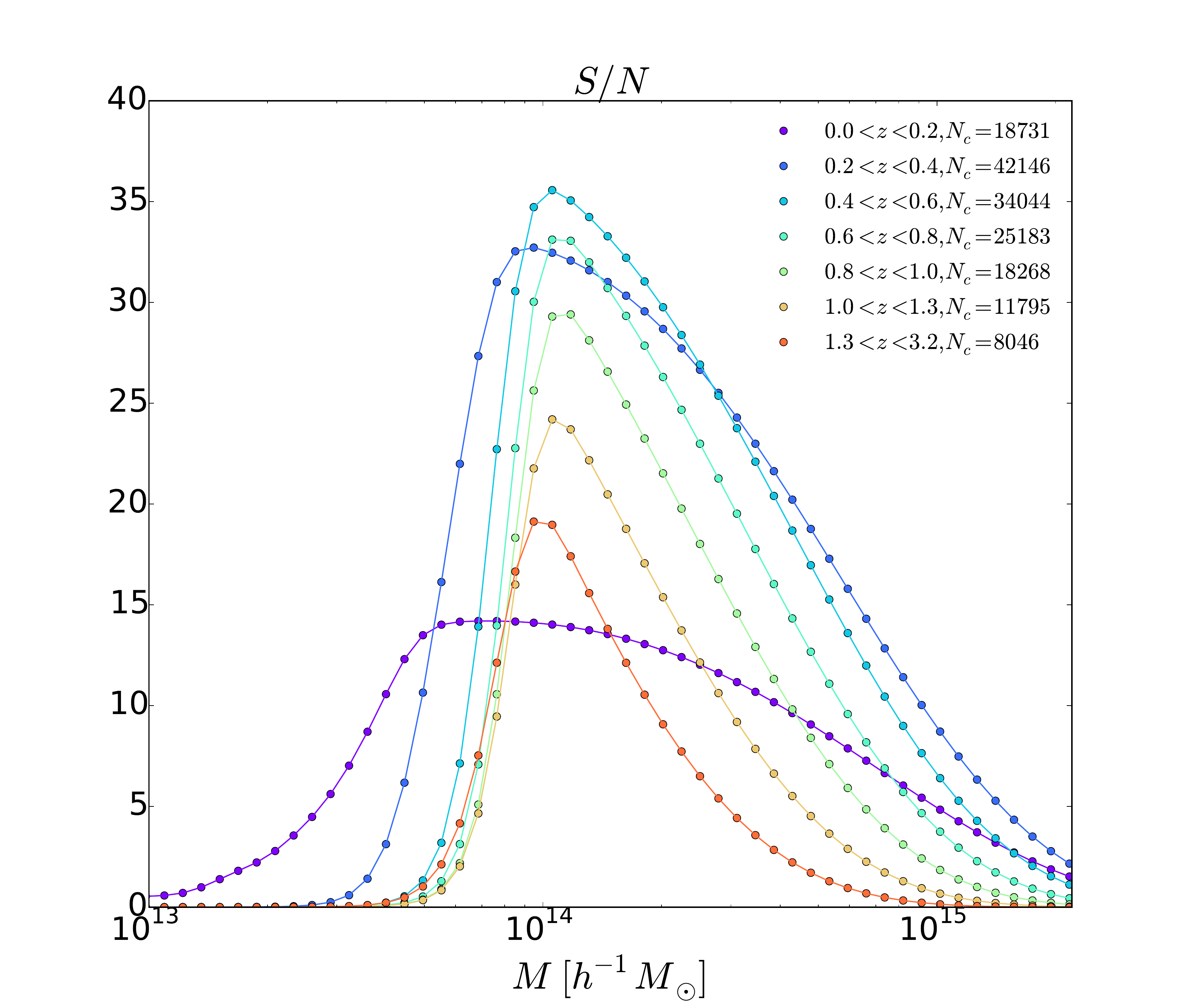}
  \caption{$S/N$ on the reconstruction of the ${ \cal Y}-M$ in seven bins
           of redshift. A S4 CMB experiment will be able to measure the
           redshift evolution of this scaling relation to good accuracy
           in a wide range of masses. The number of clusters found in each 
           redshift bin is indicated in the legend. The masses are estimated using only
           polarisation-based estimators.}
  \label{fig:blindredshift}
\end{figure}

We use a Fisher matrix formalism to predict the uncertainties on the amplitude
$A_{Y}$ and the power law index $\alpha_{Y}$. These come from three sources:
the measurement
uncertainties on the tSZ flux, the intrinsic scatter in the $Y-M$ relationship
and uncertainties on the cluster lensing mass measurement. The first two sources
of uncertainty are always subdominant compared to the later, and in this
section we will assume that they can be neglected. A joint likelihood formalism
that consistently accounts for all sources of uncertainty will be presented in
Section \ref{sec:cosmopar}. Within this approximation and assuming Gaussian
measurement errors, the form of the likelihood is simple
\begin{equation}
-\ln{\cal L} \propto \sum^{N_{c}}_{i=1}
\frac{\left[M^{obs}_{500,i}-f(Y_{500,i},A_{Y},\alpha_{Y})\right]^{2}}{2\sigma^{2}_{M_{500,i}}},
\end{equation}
where $f(Y_{500},A_Y,\alpha_Y)\equiv M_{500}(Y_{500},A_Y,\alpha_Y)$ is given in Eq.
\ref{eq:yscaling}. The Fisher matrix therefore takes a simple form:
\begin{equation}\nonumber
F_{\alpha \beta}= \int dz\frac{dV}{dz}
\int dM\frac{n(M,z)\tilde{\chi}(M,z)}{{\cal \sigma}^{2} (M,z)}
\frac{\partial f}{\partial \alpha}
\frac{\partial f}{\partial \beta},
\end{equation}
and all terms of the Fisher matrix can be computed analytically.

Fig.~\ref{fig:scaling} shows the resulting constraints for a S4 CMB
experiment where we consider the full reconstruction of the
convergence map as well as the reconstruction based on
polarisation-only data. The power-law index $\alpha_{Y}$ and the
amplitude $A_{Y}$ could be measured at the per-cent level, using only
CMB data.

\subsection{Non-parametric reconstruction of the Y-M relationship}
\label{subsec:nonparametric} 
    
The high accuracy on the calibration of the $Y-M$ scaling relation for CMB S4
experiments ($>200 \sigma$ determination of the hydrostatic bias) leads us to
investigate a non-parametric reconstruction of the $Y-M$ relation.
The main interest of this approach is the avoidance of a particular modelling
prior.

Let us start by defining the redshift-independent observable
${\cal Y}=d_A^{2}Y_{500}/E^{2/3}(z)$. We define a non-parametric relation
between ${\cal Y}$ and $M$ in $N_b$ bins of ${\cal Y}$ characterised by edges
$[{\cal Y}_n^i,{\cal Y}_n^f]\,(n\in[1,N_b])$ as:
\begin{equation}
  M({\cal Y})=\sum_n M_n\,W({\cal Y}|{\cal Y}^i_n,{\cal Y}^f_n),
\end{equation}
where $W(x|x_a,x_b)$ is a top-hat window function in the interval
$[x_a,x_b]$.

Under the assumption that the uncertainty in the lensing mass measurement
dominates over the uncertainty on ${\cal Y}$ (which receives contributions
from the measurement uncertainties as well as the intrinsic scatter), and
following the same $\chi^2$ argument used in the previous section, we can
compute the expected errors on $M_n$ as
\begin{equation}
\sigma^{-2}(M_n)=
\int dz\frac{dV}{dz} \int_{M_n^0}^{M_n^f}dM
\frac{n(M,z)\tilde{\chi}(M,z,b)}{\sigma^{2}(M_L)}.
\end{equation}
Figure \ref{fig:blind} shows the 1$\sigma$ uncertainties on this
${\cal Y}-M$ relationship for $N_b=44$ logarithmic bins of mass in the range
$M\in[2 \times 10^{13}, 2 \times 10^{15}]\,M_\odot/h$, as well as the signal-to-noise ratio in each bin.
Note that while we illustrate the method assuming a simple power-law relation
between ${\cal Y}$ and $M$, the error on each mass bin is independent of this
particular choice, and thus this method could be used to detect deviations
from this fiducial model, providing significant insight into the
physics of clusters. The $S/N$ is a function of the number of cluster detected
inside a given mass bin and the error on the reconstruction of the convergence
field. A CMB S4 experiment would be able to reconstruct the cluster $Y-M$ relation
with $S/N>5$ for 38 bins in the mass range $[2.9\times 10^{13},
1.5 \times10^{15}]M_\odot/h$, peaking at masses $\sim10^{14}  M_\odot/h$ using
measurements of the lensing signal from polarisation data only.

This method can, also be used to study the redshift dependence of the 
$Y-M$ relation. As an example, Figure \ref{fig:blindredshift} shows the possible
constraints on the non-parametric model in 7 different bins of redshift in the
range $0\leq z\leq3.2$. Significant constraints on the scaling relationship can
still be drawn in this case in a wide range of masses for all redshifts.

\section{Joint constraints from tSZ and lensing measurements}\label{sec:cosmopar}
\begin{figure*}
  \centering
  \includegraphics[width=0.8\textwidth]{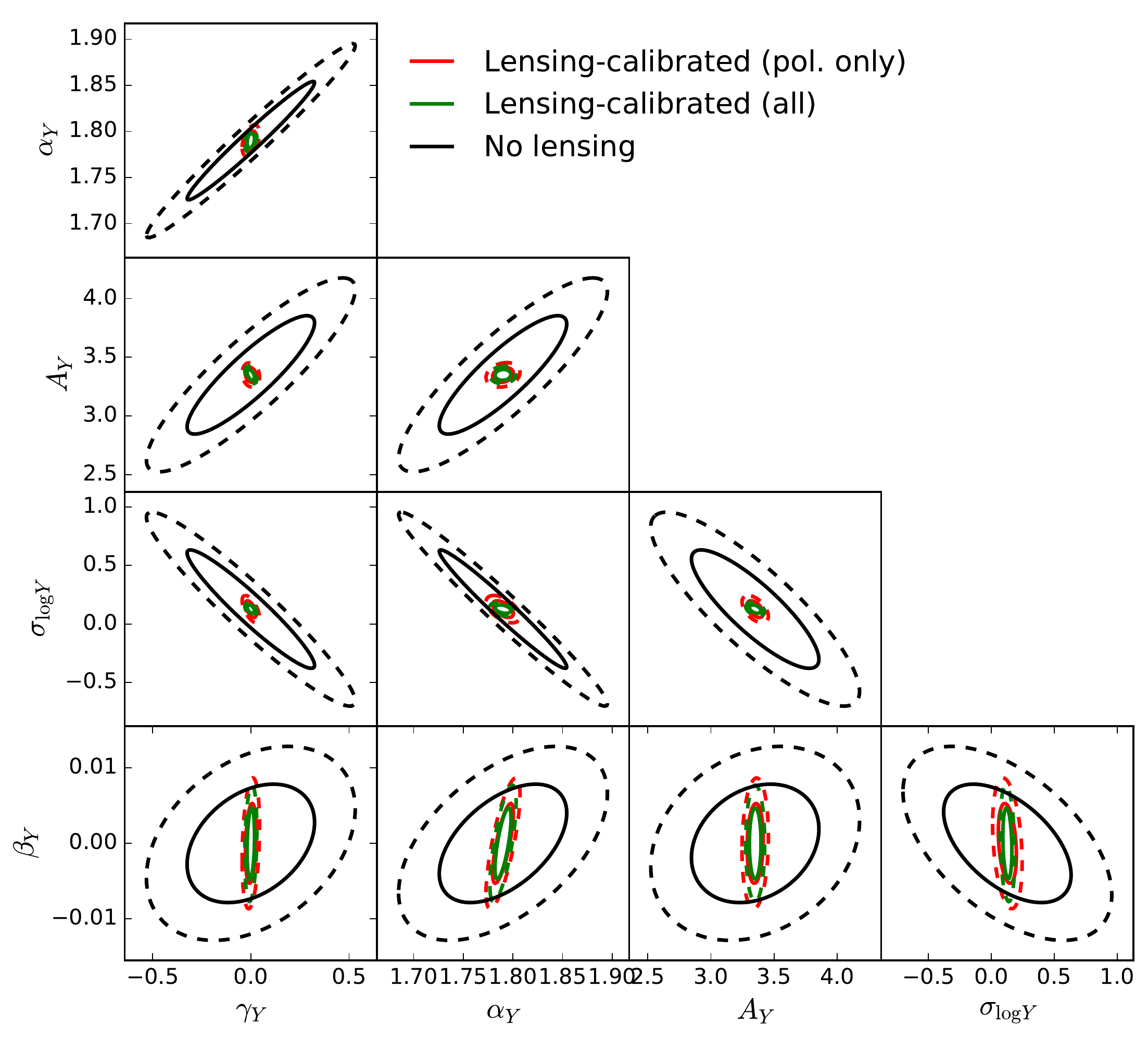}
  \caption{Uncertainties on the cluster nuisance parameters marginalised over the
           cosmological parameters for an SZ survey carried out with S4 without
           lensing information (black ellipses), with lensing masses reconstructed
           using only polarisation information (red ellipses) and using also
           temperature (green ellipses). In all cases we include prior information
           on the cosmological parameters from Planck.}
  \label{fig:nuisance_cluster}
\end{figure*}
\begin{figure}
  \centering
  \includegraphics[width=0.48\textwidth]{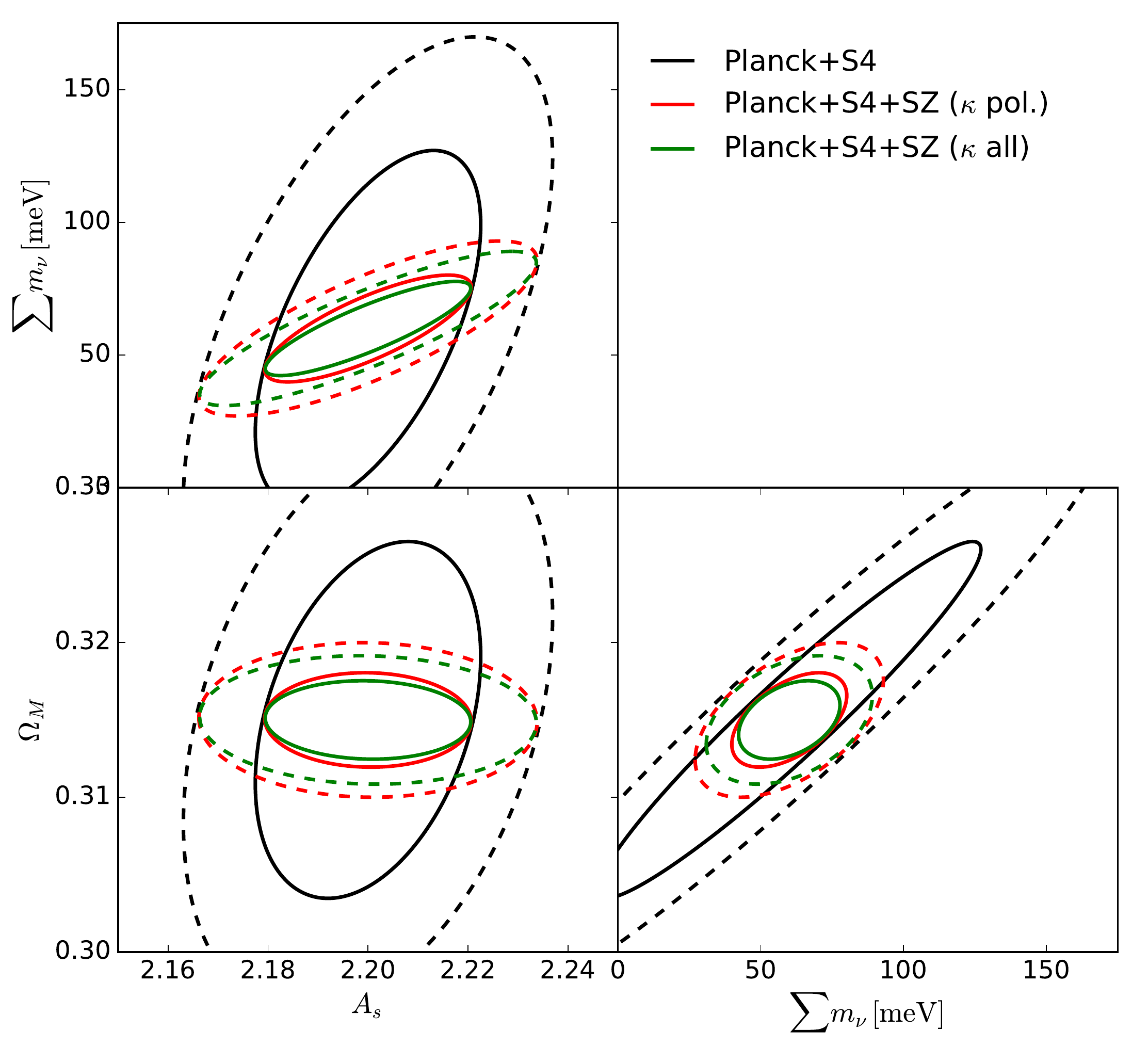}
  \caption{Uncertainties $\Omega_M$, $A_s$ and the sum of neutrino masses from 
           an SZ catalog carried out with S4 in combination with constraints from S4
           primary and lensing power spectra, as well as Planck temperature and polarisation
           on $\ell<30$. Results are shown in the absence of lensing mass estimates 
           (black ellipses), and for lensing masses computed using only polarisation
           (red ellipses) and temperature and polarisation (green ellipses). The results
           are marginalised over all other cosmological parameters as well as the cluster
           nuisance parameters.}
  \label{fig:cosmopar_cluster}
\end{figure}
\begin{figure}
  \centering
  \includegraphics[width=0.48\textwidth]{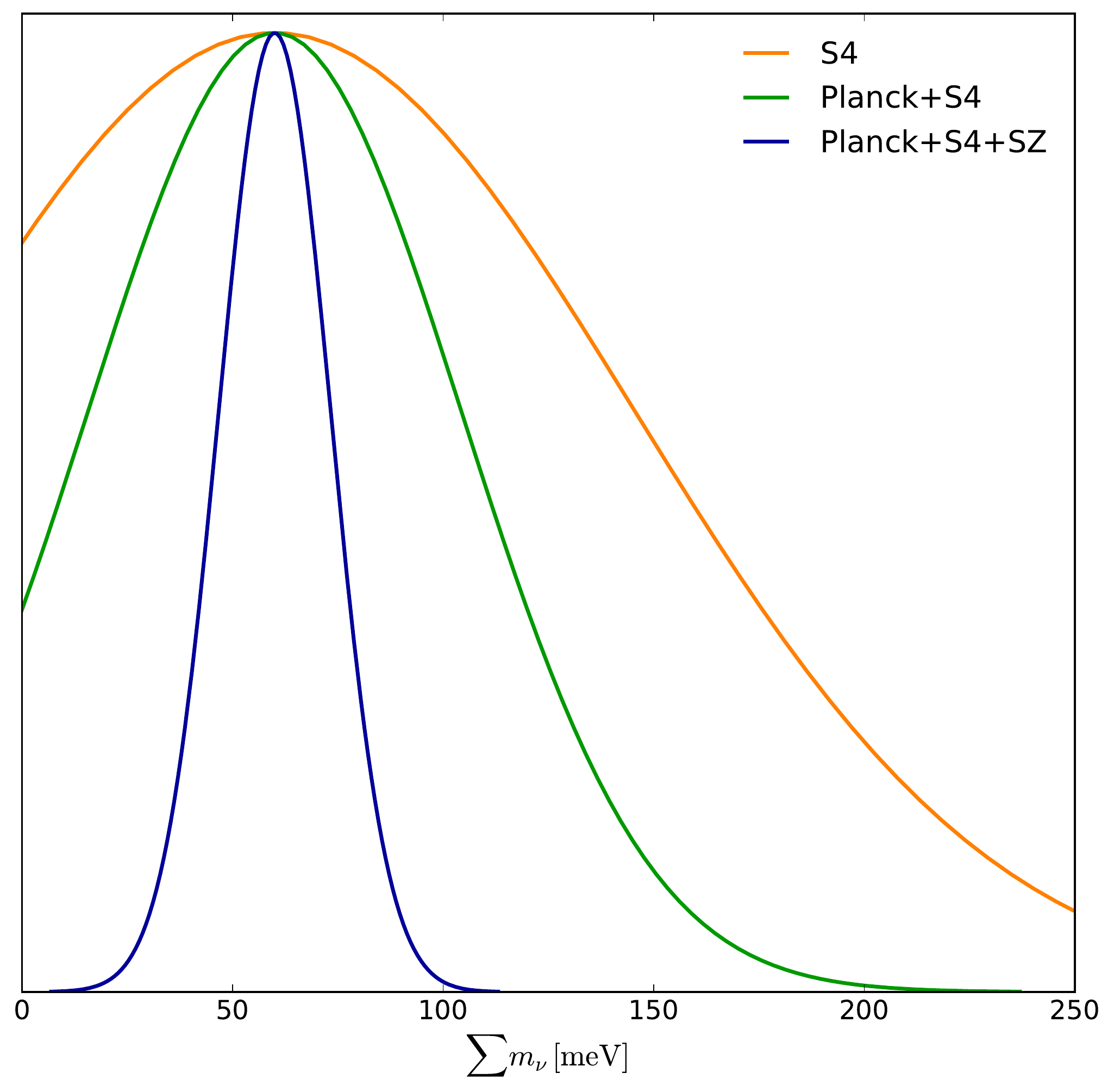}
  \caption{Posterior distribution for the sum of neutrino masses for different experiment
           configurations: S4 primary and lensing power spectra (gray curve), S4 power
           spectra and Planck primary on $\ell<30$ (orange curve), S4 and Planck power
           spectra together with SZ number counts from S4 in the absence of lensing masses
           (green curve) and for lensing masses measured using only polarisation information
           (blue curve).}
  \label{fig:mnu_cluster}
\end{figure}
The aim of this section is to provide a general formalism to consistently
account for the uncertainties in the $Y-M$ relation when drawing cosmological
constraints from cluster number counts. This is of particular interest in
the presence of lensing data which, as we have seen, can directly constrain this
relation. Assuming that we have measured the redshift $z$, tSZ flux
$Y_{\rm obs}$ and lensing mass $M_L$ for each cluster in the catalog,
we use as our basic observable the number of clusters detected in bins of these
three quantities:
\begin{equation}
  \frac{N(q,M_L,z)}{\Delta q_Y\,\Delta M_L\,\Delta z}\equiv
  \int d M d Y\frac{d^3N}{dz\,d M\,d Y}
  P(q_Y,M_L|Y,M),
\end{equation}
where $P(q_Y,M_L|Y,M)$ is the probability of measuring a lensing mass
$M_L$ and a tSZ flux with a signal-to-noise $q_Y\equiv
Y_{\rm obs}/\sigma_Y$ for a cluster of true mass $M$ and tSZ flux $Y$.
Note that we have chosen to use $q_Y$ instead of $Y_{\rm obs}$ as the
observable variable, since the fiducial cluster catalog assumed here is
defined by a threshold $q_Y>6$. We proceed by making the following
assumptions:
\begin{enumerate}
 \item The tSZ and lensing measurements are independent, and the noise on
       these quantities has a Gaussian distribution. In this case we can
       write:
       \begin{equation}
         P(q_Y,M_L|Y,M)={\cal N}(q_Y|Y/\sigma_Y,1)\,
         {\cal N}(M_L|M,\sigma_\kappa),
       \end{equation}
       where ${\cal N}(x|\mu,\sigma)$ is the normal distribution with mean
       $\mu$ and variance $\sigma^2$, and $\sigma_Y$ and $\sigma_L$ are the
       errors in the measurement of $Y_{\rm obs}$ and $M_L$.
 \item The true tSZ flux and halo mass are related through a stochastic
       log-normal model, such that $\log Y= \log \bar{Y}(M,z)+n_Y$,
       where $n_Y$ is a random normal variable with mean 0 and standard
       deviation $\sigma_{\log Y}$. $\bar{Y}(M,z|{\bf p})$ is a scaling
       relation dependent on a set of nuisance parameters ${\bf p}$ that
       we will specify later on. We can then write
       \begin{equation}
         \frac{d^3N}{dz\,dM\,d\log Y}=\frac{dN^2}{dz\,dM}\,
         {\cal N}(\log Y|\log\bar{Y},\sigma_{\log Y}),
       \end{equation}
       where
       \begin{equation}
         \frac{d^2N}{dzdM}=4\pi f_{\rm sky}\,\frac{c\,r^2(z)}{H(z)}\,n(M,z)
       \end{equation}
       and $n(M,z)$ is the halo mass function.
 \item Finally, we assume that the fluctuations in the counts of objects in
       different bins of $q_Y$, $z$ and $M_L$ are independent and
       Poisson-distributed, such that the likelihood for a given set of
       counts $N_{\rm obs}$ is:
       \begin{equation}\label{eq:likefull}
         \ln{\cal L}(N_{\rm obs}|\bar{N})=
         \sum_{z,q_Y,M_L}\left[N_{\rm obs}\,\ln\bar{N}-\bar{N}-
         \ln(N_{\rm obs}!)\right]_{z,q_Y,M_L},
       \end{equation}
       where $\bar{N}(z,q_Y,M_L)$ is the mean number of clusters
       given above, which depends both on the cosmological parameters
       and the $Y-M$ relation. 
\end{enumerate}

Expanding Eq. \ref{eq:likefull} around the maximum likelihood point, we find the
expression for the Fisher matrix \cite{2011MNRAS.412.1895S}:
\begin{equation}
  F_{\alpha\beta}=\sum_{z,q_Y,M_L}
  \frac{\partial_\alpha\bar{N}(z,q_Y,M_L)\,
  \partial_\beta\bar{N}(z,q_Y,M_L)}
  {\bar{N}(z,q_Y,M_L)}.
\end{equation}
We can then produce forecasts for cosmological constraints from number counts
consistently marginalised over the $Y-M$ relation by computing the Fisher matrix
above including both the cosmological parameters and the mass calibration parameters ${\bf p}$.
For this analysis we will use a generalised version of the simple scaling relation
\ref{eq:yscaling} to account for additional variation with respect to mass and redshift:
\begin{align}\label{eq:yscaling_generic}\nonumber
 E^{-2/3}(z)&\left[\frac{d_A}{100\,{\rm Mpc}/h}\right]^2\,
 \frac{\bar{Y}(M,z)}{10^{-10}\,{\rm srad}}=\\
 &A_Y\left(\frac{M}{M_*}\right)^{\alpha_Y}e^{\beta_Y\log^2(M/M_*)}
 (1+z)^{\gamma_Y}.
\end{align}
The set of nuisance parameters for this model is therefore
${\bf p}\equiv(A_Y,\alpha_Y,\beta_Y,\gamma_Y,\sigma_{\log Y})$, for which we will use
the fiducial values reported in \cite{2015arXiv150201597P}:
${\bf p}=(3.35\times10^{-10}{\rm srad}^2,1.79,0,0,0.127)$ (for a pivot scale
$M_*=1.5\times10^{14}\,M_\odot\,h^{-1}$).

We produce constraints for a cluster sample divided into 64 logarithmic bins of $q_Y$
and $M_L$ in the ranges $q_Y\in[6,500]$,
$\log_{10}M_\kappa/(M_\odot\,h^{-1})\in[12.5,15.5]$, and 10 bins of
redshift between $z=0$ and $z=2$. Note that in principle it should be possible to use
much narrower redshift intervals, however we choose to use wide bins ($\Delta z=0.2$) to
justify using a purely Poisson likelihood, ignoring the sample covariance caused by
the average fluctuation of the density field inside each bin \cite{2003ApJ...584..702H}.

In order to include in our forecasts the cosmological constraints
achievable with other complementary probes, we have also carried out
a Fisher matrix forecast for measurements of the CMB primary and lensing power spectra
using the formalism described in \cite{2015ApJ...814..145A}. In this case we have considered
two experimental setups, corresponding to S4 as described in Table \ref{tab:cmbexp}, and
to the Planck experiments \cite{2014A&A...571A...1P}, modelled assuming a map-level RMS noise
of $43\uKam$, a Gaussian beam of 7 arcmin FWHM and a sky fraction $f_{\rm sky}=0.7$. In both
cases we impose a maximum $\ell$ cut of $\ell_{\rm max}=3000$ in intensity and
$\ell_{\rm max}=5000$ in polarisation, and in the case of S4 we assume a minimum $\ell$ cut
of $\ell_{\rm min}=30$. When combining both
experiments we assume S4-only and Planck-only constraints on 40\% and 30\% of the sky
respectively for $\ell>30$, and Planck-only constraints on 70\% of the sky for $2<\ell<30$. The resulting
Fisher matrix is then directly added to the Fisher matrix for cluster number counts computed
as described above, under the assumption that the constraints from both probes are
uncorrelated. In all cases we marginalise over a set of 7 cosmological parameters:
the matter density parameter $\Omega_m$, the baryon fraction $f_b\equiv\Omega_b/\Omega_M$,
the Hubble parameter $h$, the scalar spectral index $n_s$ and amplitude $A_s$, the
optical depth $\tau$ and the sum of neutrino masses $\sum m_\nu$. For these parameters we
use the fiducial values mentioned in Section \ref{sec:intro}, as well as the  value
$\sum m_\nu=60\,{\rm meV}$, corresponding to the lower limit allowed by neutrino oscillation
experiments. Together with the cluster nuisance parameters introduced above we therefore consider
a 12-dimensional parameter space. Power spectra were computed using the public code {\tt CLASS}
\cite{2011JCAP...07..034B}.

The red and green ellipses in Figure \ref{fig:nuisance_cluster} show the constraints on
the nuisance parameters of the scaling relation for lensing mass measurements carried out
using polarisation-only quadratic estimators, and all estimators respectively. The constraints
are marginalised over all cosmological parameters, and show the small effect of discarding
temperature data on the cluster parameters. It is worth pointing out
that, even in the absence of mass measurements it would be possible to constrain the
parameters of the $Y-M$ relation to some extent, since they affect the observed cluster
mass distribution, which is independently constrained from N-body simulations. The
achievable constraints in
the absence of cluster lensing masses are represented by the black ellipses in the same
figure. The effect of cluster lensing information on the constraints for the most relevant
cosmological parameters is shown in Figure \ref{fig:cosmopar_cluster} using the same
color code. Of particular relevance is the factor $\sim3.3$ improvement in the uncertainty
on the sum of neutrino masses, a key science case for S4. We further showcase the role of
cluster number counts in constraining this parameter in Figure \ref{fig:mnu_cluster}.
While the constraints on $\sum m_\nu$ from cluster abundances alone are irrelevant
given current bounds, and achievable constraints with S4 in combination with Planck 
using only power spectrum information would not yield a significant measurement of
neutrino masses, the combination of both probes can efficiently break degeneracies
between different cosmological parameters, enabling a
$\sim5\sigma$ measurement of $\sum m_\nu$ ($\sigma(\sum m_\nu)=13.2\,{\rm meV}$)
when using lensing mass information to constrain the $Y-M$ relation.

\section{Discussion}\label{sec:conclusion} 
We have studied the potential of using CMB lensing by clusters to calibrate their
masses in the era of CMB Stage-4 experiments. We have found that CMB S4 will allow
a sub-percent determination of the hydrostatic bias parameter relating the mass 
inferred from X-ray observation to the true cluster mass. We have then extended 
the model and shown that the large number of detected clusters and the low
uncertainties in the reconstructed convergence maps could be used to calibrate
the $Y-M$ relationship using solely CMB data. We have also studied the constraints
on a non-parametric reconstruction of the $Y-M$ relationship which can be used to
study the evolution of this relation with mass and redshift with high 
significance over a wide range of masses. Throughout the paper we have studied
the impact of discarding temperature data in the reconstruction of the
convergence field. This is an important comparison, since the temperature data
from Stage-4 CMB experiments will suffer from atmospheric contamination and
residual foregrounds. Finally we have presented a joint likelihood for tSZ and
lensing mass measurements allowing us to forecast constraints on cosmological
parameters while consistently accounting for the uncertainties in the $Y-M$
scaling relation.

The method presented here relies on a number of assumptions. Testing and
characterising the effect of each of these assumptions is the subject of
future work but it is worth quoting the following caveats, which might affect
future analyses with real data:
 \begin{itemize}
 \item Throughout this paper, we have modelled the lensing field and the
       lensing reconstruction noise as Gaussian fields. At the level of
       precision  achieved by CMB Stage-4 experiments this assumption
       may break down \cite{2012PhRvD..86l3008B,2016arXiv160803169L}, and
       the non-Gaussian contribution to the signal and/or the noise should
       be studied in detail.
 \item We used a Poisson likelihood for cluster number counts. While this
       is commonly used for current cluster catalogs, the high number of
       clusters detected in the CMB S4 catalog might require a more
       sophisticated treatment. Studies of possible departure from Poisson
       for cluster number count can be found in \cite{2003ApJ...584..702H,
       2004PhRvD..70d3504L, 2011MNRAS.418..729S}, where a sample variance
       term is added to the Poisson shot-noise term to account for the fact
       that cluster are peaks of the same underlying density field. In our
       analysis we have used wide redshift bins to reduce the impact of sample
       variance.
 \item We have not included the covariance between the CMB lensing power
       spectrum and cluster counts calibrated using CMB lensing when 
       constraining cosmological parameters. In practice, the full
       covariance could be constructed from Monte-Carlo simulations while
       analysing CMB S4 data (e.g. see \cite{2016arXiv160105779K}).
 \item The cluster mass function used to determine cosmological parameters
       is a fit to N-body simulations \cite{2008ApJ...688..709T}. In order
       to achieve the accuracy required for S4, in particular for
       non-standard scenarios such as massive neutrinos \cite{2013JCAP...12..012C},
       a better understanding of the theoretical uncertainties in the mass
       function is necessary, in particular regarding the effects of baryonic
       physics \cite{2012MNRAS.423.2279C}. 
 \item Finally, we have assumed knowledge of the cluster mass
       \cite{1996ApJ...462..563N} and pressure \cite{2010A&A...517A..92A}
       profiles, and this assumption allows us to devise a minimum-variance
       estimate of $Y$ and $M$ from the data. However, this needs to be
       validated using hydrodynamic simulations at the level of precision
       corresponding to the uncertainties on cosmological parameters. A more
       careful analysis of cluster de-blending is also necessary
\end{itemize}
While more work need to be done to address each of these issues, we have shown that
using cluster lensing to calibrate cluster masses has an important potential and
could play a role in the future determination of the sum of neutrinos masses.
This, combined with particle physics measurements, could allow us to distinguish
between the normal and inverted hierarchies, thus opening a new window on
fundamental physics.

\section*{Acknowledgments}
 We thank Nicholas Battaglia, Joanna Dunkley, Pedro Ferreira, Mathew Madhavacheril,
 Sigurd N{\ae}ss and  Joseph Silk, for useful comments and discussions. TL is
 supported by ERC grant 267117 (DARK) hosted by Universite Pierre et Marie
 Curie- Paris 6 and by the Labex ILP (reference ANR-10-LABX-63) part of the Idex
 SUPER, and received financial state aid managed by the Agence Nationale de la
 Recherche, as part of the programme Investissements d'avenir under the
 reference ANR-11-IDEX-0004-02. DA is supported by the Beecroft Trust and
 ERC grant 259505.

\bibliography{paper}

\appendix

\end{document}